\documentclass{article}

\PassOptionsToPackage{numbers, compress}{natbib}

\newcommand{\boldres}[1]{{\textbf{\textcolor{red}{#1}}}}
\newcommand{\secondres}[1]{{\underline{\textcolor{blue}{#1}}}}

\newcommand{\revision}[1]{{\textcolor{black}{#1}}}
\newcommand{\method}{\textsc{MiFormer}\xspace}

\usepackage[preprint]{neurips_2025}

\usepackage[utf8]{inputenc} 
\usepackage[T1]{fontenc}    
\usepackage{hyperref}       
\usepackage{url}            
\usepackage{booktabs}       
\usepackage{amsfonts}       
\usepackage{nicefrac}       
\usepackage{microtype}      
\usepackage{xcolor}         

\usepackage[pdftex]{graphicx}
\usepackage{amsmath}
\usepackage{multirow}
\usepackage{xspace}

\DeclareMathOperator*{\argmin}{argmin}
\DeclareMathOperator*{\argmax}{argmax}

\title{UDuo: Universal Dual Optimization Framework for Online Matching}

\author{%
  Bin Li \quad Diwei Liu \quad Zehong Hu \quad Jia Jia\\
  Alibaba Group \\
  Hangzhou \\
  \texttt{{binli.lb}@hotmail.com} \\
  \texttt{\{zehong.hzh,jj229618\}@alibaba-inc.com} \\
  \texttt{{diwei.ldw}@koubei.com}
}

\begin{document}

\maketitle

\begin{abstract}
Online resource allocation under budget constraints critically depends on proper modeling of user arrival dynamics. Classical approaches employ stochastic user arrival models to derive near-optimal solutions through fractional matching formulations of exposed users for downstream allocation tasks. However, this is no longer a reasonable assumption when the environment changes dynamically. In this work, We propose the \textbf{U}niversal \textbf{Du}al \textbf{o}ptimization framework (\textbf{UDuo}), a novel paradigm that fundamentally rethinks online allocation through three key innovations: (i) a temporal user arrival representation vector that explicitly captures distribution shifts in user arrival patterns and resource consumption dynamics,
(ii) a resource pacing learner with adaptive allocation policies that generalize to heterogeneous constraint scenarios, and (iii) an online time-series forecasting approach for future user arrival distributions that achieves asymptotically optimal solutions with constraint feasibility guarantees in dynamic environments. Experimental results show that UDuo achieves higher efficiency and faster convergence than the traditional stochastic arrival model in real-world pricing while maintaining rigorous theoretical validity for general online allocation problems.
\end{abstract}

\section{Introduction}
Online matching has become a fundamental problem in real-world resource allocation systems, with critical applications spanning dynamic pricing, bid optimization, and internet advertising. When complete user distribution information is available, such problems are reduced to offline linear programming formulations whose optimal solutions serve as theoretical upper bounds for online counterparts where decisions must be made without future visibility. 

To address the inherent uncertainty in user arrivals, previous works have developed two dominant modeling paradigms: adversarial models assuming malicious input sequences crafted by omniscient adversaries while typically over-pessimistic in practice, and stochastic models positing stationary user distributions. Most studies rely on stochastic models and employ primal-dual algorithms to compute fractional matching problems from historical data, generating dual solutions that guide downstream allocations.
These methods achieve near-optimal solutions while under stationary environments. However, practical applications often operate in dynamically changing environments where relying on static optimal dual solutions for neighboring users is inherently inadequate. Zhou et al. \cite{zhou2019robust} proposed a novel user arrival model assuming users are drawn from a drifting distribution. While this approach performs better under user sequences with abrupt peaks, it does not necessarily lead to improved performance when environments evolve gradually.

Our key contributions include: (i) We introduce a novel user arrival representation method and mathematically analyze its expressive capability for characterizing user distributions; (ii) We propose UDuo, a general-purpose online matching dual optimization framework equipped with time-series forecasting capabilities, theoretically proving its constraint feasibility guarantees; (iii) We develop a pacing module that enables fine-grained temporal resource allocation through temporal-aware or generative approaches; (iv) We design a \textbf{M}ulti-scene \textbf{i}Trans\textbf{Former} (\textbf{MiFormer}) to support generalizable time-series forecasting capabilities for solver modules across diverse response models and traffic patterns. Experimental validation on dynamic pricing in real-world food delivery scenarios demonstrates that UDuo achieves higher efficiency and faster convergence compared to conventional stochastic arrival model-based online optimization methods. Furthermore, MiFormer reduces average squared errors by 8\% across four benchmark datasets and two production-scale datasets collected from food delivery platforms, outperforming state-of-the-art time series foundation models.

\section{Preliminaries}
\label{preliminaries}

For simplicity and generality in theoretical analysis, we define the online resource allocation framework as follows. Let user $i \in  \mathcal{C} = \{1,...,|\mathcal{C}|\}$ where $\mathcal{C}$ denote the set of users, and \( \mathcal{T} = \{1,...,|\mathcal{T}|\}\) represent the treatment candidates where \(j\in \mathcal{T} \) corresponds to the \(j\)-th incentive item. The total budget \( \mathcal{B} \in\mathbb{R}_{>0} \) serves as the global resource constraint. At each round $t$, the $i$-th user receives a reward response $r_{ij} \in \mathbb{R}^\mathcal{T}$ specific to treatment $j$ from the uplift model. And $c_{ij} \in \mathbb{R}^\mathcal{T}$ quantifies the cost of assigning treatment $j$ to user $i$. The online optimizer selects a decision vector $x_{ij}\in\{0,1\}$ where $x_{ij}$ equals 1 to indicate assigning treatment $j$ to user $i$. Then we relax the integer programming to formulate the problem as a linear programming (LP) problem. Formally,

\begin{equation}
    \begin{aligned}
\max_{x_{ij}} &\sum_{i=1}^{|\mathcal{C}|} \sum_{j=1}^{|\mathcal{T}|}  x_{ij}r_{i j} \\
s.t.\
&\sum_{j=1}^{|\mathcal{T}|} x_{ij} \leq 1,for\ i=1,...,|\mathcal{C}|\\
&\sum_{i=1}^{|\mathcal{C}|} \sum_{j=1}^{|\mathcal{T}|} x_{i j} c_{ij}  \leq \mathcal{B}
\end{aligned}
\label{obj_eq}
\end{equation}

We apply the Lagrange multiplier method to transform the primal problem into its dual counterpart:
\begin{equation}
\begin{aligned}
\label{dual-obj}
&\min_{\lambda\geq0}(\lambda \mathcal{B}+\sum_{i=1}^{|\mathcal{C}|}\sum_{j=1}^{|\mathcal{T}|}x_{ij}(r_{i j} - \lambda c_{ij})) \\
=&\min_{\lambda\geq0}(\lambda \mathcal{B}+\sum_{i=1}^{|\mathcal{C}|}\max_{j} (r_{i j} - \lambda c_{ij}))
\end{aligned}
\end{equation}

where $\lambda \in \mathbb{R}_{>0}$ is the dual variable and the optimal $\lambda^*$ can be obtained using the gradient descent algorithm L-BFGS~\cite{liu1989limited} or the binary search~\cite{zhou2023direct}. The optimal decision can be derived as follows.

\begin{equation}
\label{rank_eq}
\begin{aligned}
    x_{ij}= \begin{cases}1, & \text { if } j=\argmax_{j} r_{ij}-\lambda^*  c_{ij} \\ 0, & \text { otherwise }\end{cases}
\end{aligned}
\end{equation}
During online matching decisions, we utilize the real-time computed value of $\lambda^*$ along with decision Eq~\ref{rank_eq} to price the incentive value.

\begin{figure*}[t]
\centering
\includegraphics[width=0.96\linewidth]{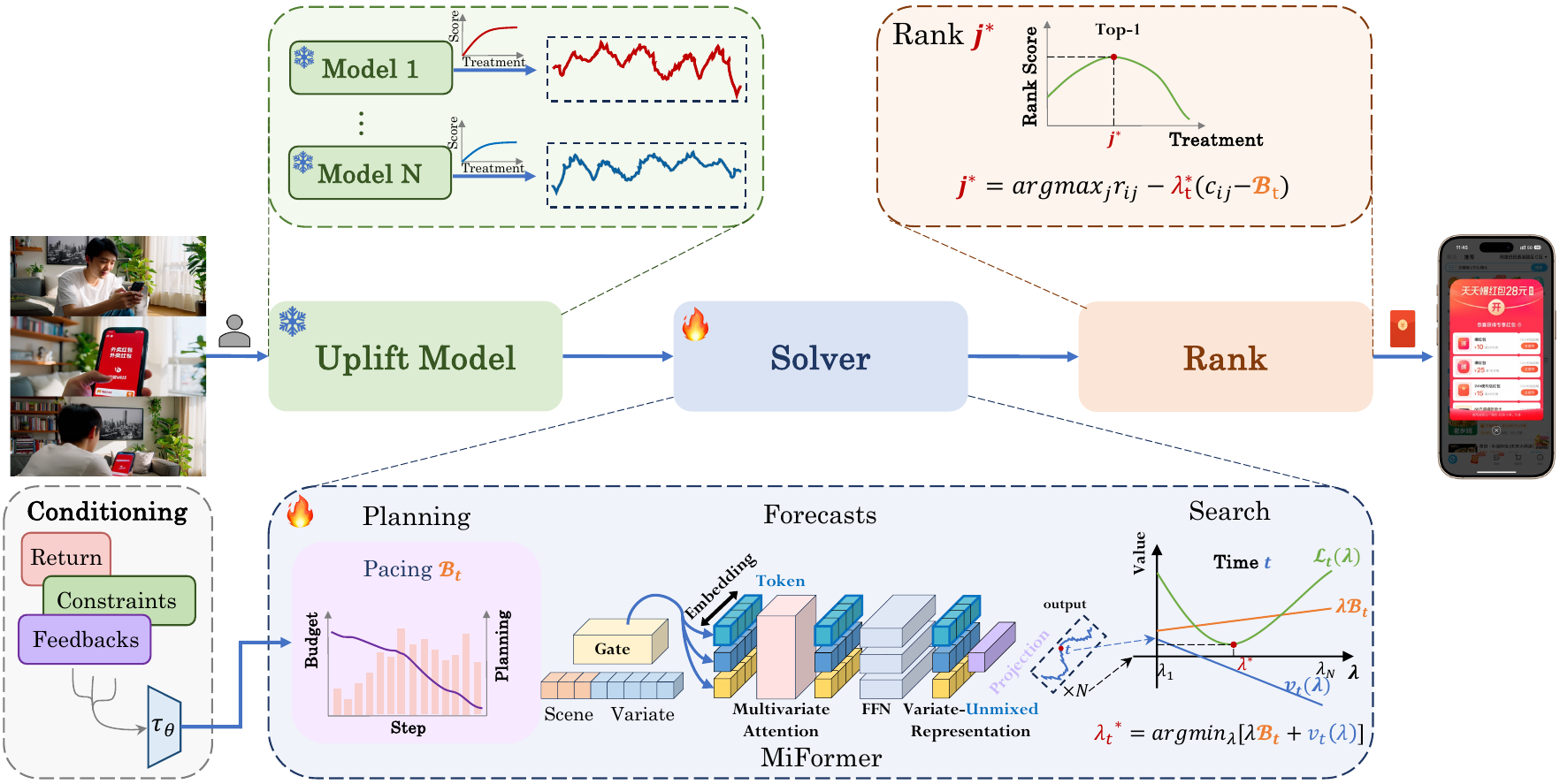}
\caption{Illustration of the \textbf{UDuo} framework. Take the food delivery scenario as an example: When a user initiates a coupon redemption request, we first apply a pretrained uplift model with frozen parameters to generate response scores for multi-treatment coupon denominations, capturing each user's incremental response propensity to different discounts. Subsequently, the target user cohort is aggregated into temporal sequences of user arrival representation vectors via Eq.~\ref{vlambda}. Next, our online solver module performs a binary search over future time slots using pacing-adjusted budgets and predicted $v(\lambda)$ vectors to derive optimal dual solutions. Finally, these dual solutions feed into the decision Eq.~\ref{rank_eq} to finalize the ranking and allocation of coupon denominations.}

\vspace{-1em}
\label{fig:solver}
\end{figure*}

\section{Methodology}
\label{method}

\subsection{User Arrival Modeling}

The user arrival model focuses on two critical aspects: i) quantifying the expected number of future users and their dynamic distribution patterns. ii) measuring users' sensitivity to allocated incentives (e.g., coupons) and associated resource consumption. Through analysis of the dual objective Eq~\ref{dual-obj}, we identify four core components in its second term that collectively characterize user arrival dynamics: the incremental response scores $r_{ij}$ generated by uplift models to quantify users' sensitivity to treatment $j$, resource consumption costs $c_{ij}$ representing the expenditure of assigning treatment $j$ to user $i$, the dual variable $\lambda$ reflecting resource scarcity, and the population scale factor $|\mathcal{C}|$ indicating total user volume. Furthermore, we define the user arrival representation vector $v(\lambda)$ as follows:
\begin{equation}
\label{vlambda}
    v(\lambda)= \sum_{i=1}^{|\mathcal{C}|}\max_{j}(r_{i j} -\lambda c_{ij})
\end{equation}
The dual problem is transformed into:
\begin{equation}
    L(\lambda)=\min_{\lambda\geq0}\lambda \mathcal{B}+v(\lambda)
\end{equation}
Through the aforementioned representation vector process, we effectively encode user distribution patterns, demand intensity, and cost-effectiveness characteristics within online allocation decisions. Formally, we discretize the user arrival dynamics into temporal slices along the one-day horizon $T$. For each time period $t\in T$, $\mathcal{B}_t$ denotes the budget allocated for period $t$, and $v_t(\lambda)$ represents the temporal user representation sequence conditioned on $\lambda$, where $V(\lambda)=\{v_t(\lambda)\}_{t=1}^{T}$. The formulation enables proactive decision-making by forecasting future user representation vectors $v_{t+H}(\lambda)$ and corresponding budget allocations $\mathcal{B}_{t+H}$, with systematic coordination performed through pacing mechanisms to guarantee global budget feasibility ($\sum_{t=1}^{T}\mathcal{B}_t\leq \mathcal{B}$) while optimizing resource efficiency.

To circumvent the limitations of stationary i.i.d. assumptions in previous stochastic models, we propose the Universal Dual Optimization Framework (UDuo), a generalizable online dual optimization paradigm that leverages temporal user arrival representation vector, as shown in Fig.~\ref{fig:solver}. Unlike conventional approaches relying on primal-dual algorithms over historical user distributions from stochastic arrival models, UDuo predicts future user distribution representations $v_{t+H}(\lambda)$ via time-series modeling, replacing the neighboring optimal solution approximation. Deriving optimal dual solutions $\lambda_{t+H}$ under dynamic constraints through budget pacing mechanisms. Achieving equivalent optimality to gradient descent with $\mathcal{O}(log|\mathcal{C}|\frac{1}{\epsilon})$ complexity via binary search, where $\epsilon$ precision is controlled by discretization granularity. This formulation enables real-time matching decisions while maintaining rigorous theoretical guarantees on budget feasibility and regret bounds.

\subsection{Budget Pacing}
As established in Section~\ref{preliminaries} and Eq.~\ref{vlambda}, the user arrival representation vector $v(\lambda)$ characterizes user distribution dynamics independent of resource constraints. To enhance system robustness and generalization across heterogeneous scenarios, we introduce the pacing component in the solver module and formalize two budget pacing strategies.
\paragraph{Temporal-Aware Pacing}
This strategy leverages domain-specific temporal patterns derived from historical data analysis. In food delivery applications, user traffic exhibits strong periodicity with peak demands during meal periods. Similarly, coffee purchases demonstrate diurnal patterns concentrated in morning and afternoon tea hours. We determine the periodicity of the consumption rate by using a Fourier analysis of historical traffic, which we use to adjust the rhythm of budget allocation.
\paragraph{Generative Pacing}
This approach formulates pacing as a conditional sequence generation problem  $\{\mathcal{B}_t\}_{t=1}^T~p(\dot|B,R,P)$, where $\mathcal{B}$ represents constraint target (total budget resource), $\mathcal{R}$ denotes returns specifications (expected ROI targets), and $\mathcal{P}$ encodes prior knowledge feedbacks (e.g., holiday effects). The generative model produces temporally coherent pacing sequences by condition input.

Both strategies demonstrate superior allocation efficiency compared to uniform pacing based on the stochastic arrival model while satisfying the same total budget constraint.

\subsection{Time-series Forecasting}
The future user distribution patterns $v_{t+H}(\lambda)$ are predicted using an adaptive sliding temporal window $W_t={v_{t-L}(\lambda),...,v_t(\lambda)}$ of fixed backcast length $L$, where historical sequences dynamically update with current timestamp $t$. For discrete input $\lambda$ , distinct temporal forecasts are generated to support decision-making. The optimal dual variable $\lambda_t^*$ at time $t$ is determined by:

\begin{equation}
\label{ranking_eq}
    \lambda_t^*=\argmin_{\lambda_i\in\Lambda}[ \lambda_i \mathcal{B}+v_t(\lambda_i)],\ \ \ \ \Lambda=\{\lambda^{low}+k\epsilon|k=0,1,...,K\}
\end{equation}
where $\lambda^{low}=\lambda^{*'}-\frac{K}{2}\epsilon$, $\lambda^{*'}$ is solved by stochastic arrival model. Conventional time-series forecasting approaches require scenario-specific training for specialized performance, where transferring submodels trained on particular temporal distributions to distinct data regimes inevitably causes performance degradation. Although maintaining multiple scenario-specific submodels could mitigate this limitation, such strategies incur substantial operational overhead that fundamentally constrains generalization across heterogeneous scenarios.
Inspired by multi-scenario modeling paradigms~\cite{chang2023pepnet}, we propose the Multi-scene iTransformer (MiFormer) to enhance temporal adaptability through dynamic scenario-aware parameterization. The framework explicitly incorporates scene-specific prior information into temporal embeddings via a gated expansion mechanism: Given scenario identifiers (e.g., city/population cluster IDs) and corresponding dual variables $\lambda$, MiFormer constructs scene embeddings $e_s \in \mathbb{R}^k$ that are concatenated with frozen time variables embeddings $e_v \in \mathbb{R}^e$ . A dedicated gating network then performs scenario-conditioned transformation, so as to align the corresponding numerical statistical characteristics and sequence trends for different uplift models in different scenarios.

\begin{equation}
    h_v = \gamma*Gate(Concat([e_s, frozen(e_v)])) \odot e_v
\end{equation}

where $\gamma$ is the scaling factor that is set as 1.5. And $\odot$ denotes the element-wise product.

\subsection{Search Ranking}
After forecasting the discretized $v(\lambda)$ sequences, we obtain $K$ future $T$-step user arrival representation trajectories $\{v_t(\lambda)\}_{t=1}^{T}$. The optimal dual solutions $\lambda_t^*\in \Lambda (t\in T)$ are then determined through a binary search on the dual objective Eq.~\ref{ranking_eq}. The computed $\lambda_t^*$ is propagated to the rank module, where Eq.~\ref{rank_eq} determines the optimal treatment $j^*$ for the current user $i$. This completes the decision on the reward-cost balancer for real-time incentive allocation. In particular, the framework adapts to distribution shifts through a sliding-window update mechanism. At each timestamp $t$, the predictor recalibrates future dual solutions based on updated user representation sequences. This closed-loop design enables proactive adjustment to nonstationary user arrival patterns while preserving constraint satisfaction.

\section{Main Results}
\label{results}

We conducted online A/B tests on Eleme, one of China's largest food delivery platforms, to evaluate our proposed framework. All algorithms operate under identical total budget constraints to ensure fair comparison. The baseline adopts a stochastic arrival model with online gradient descent (OGD) optimization as the control group, and online binary search (OBS) provides a streamlined approach for efficient problem solving while guaranteeing equivalent solutions. As shown in Table~\ref{tab:online_metrics}, UDuo shows statistically significant improvements in all key metrics, attributed to its substantial enhancement in allocation efficiency.

\begin{table}[!htbp]
  \caption{Online Business Performance Metrics. In comparison to the online OGD method, the UDuo framework demonstrates significant improvement rates in Elme's online resource allocation.}
  \label{tab:online_metrics}
  \centering
  \begin{tabular}{lcccccc}
  \toprule
    \textbf{Methods} & \textbf{Orders} & \textbf{GMV} & \textbf{Profit}\\
    \midrule
    OBS &	+0.14\% &	+1.02\% &	+2.52\% \\
    UDuo &	+0.72\% &+3.63\% & +3.37\% \\
    \bottomrule
  \end{tabular}
\end{table}

To validate robust generalization capability of MiFormer across diverse traffic patterns, we conduct extensive experiments on four established benchmarks and two proprietary datasets from Eleme's production systems, collectively termed the User Arrival Vector (UAV) benchmark. 
As detailed results are shown in Table~\ref{tab:ltf}, MiFormer achieves remarkable performance with a 8\% reduction in mean squared error compared to competitive models trained on a specific scene. This validates the capability of simple scenario information fusion and time-varying embeddings to capture cross-scenario temporal dependencies. Detailed implementation and benchmark settings are provided in Appendix~\ref{implement}, \ref{exp_setting}.
\begin{table}[h]
\renewcommand\arraystretch{1.2}
\caption{\revision{Long-term forecasting results. 
All results are averaged from four different forecasting horizons:  $H \in \{12, 24, 48, 96\}$ for UAV and  $\{96, 192, 336, 720\}$ for the others. A lower value indicates better performance. {\boldres{Red}}: the best, \secondres{Blue}: the second best. Our full results are in.}}
\label{tab:ltf}
\vspace{-3mm}
\begin{center}
\scalebox{0.60}{
\setlength\tabcolsep{4pt}
\begin{tabular}{c|cc|cc|cc|cc|cc|cc|cc|cc|cc|cc}
\toprule

\multicolumn{1}{c|}{\multirow{2}{*}{Methods}}
&\multicolumn{2}{c|}{\textbf{\method}} &\multicolumn{2}{c|}{iTransformer} &\multicolumn{2}{c|}{DLinear} &\multicolumn{2}{c|}{PatchTST} &\multicolumn{2}{c|}{TimesNet }&\multicolumn{2}{c|}{FEDformer} &\multicolumn{2}{c|}{Autoformer} &\multicolumn{2}{c|}{LightTS} &\multicolumn{2}{c|}{Informer} &\multicolumn{2}{c}{Reformer} \\

\multicolumn{1}{c|}{} & \multicolumn{2}{c}{\scalebox{0.99}{(\textbf{Ours})}} & 
\multicolumn{2}{|c|}{\scalebox{0.99}{\citeyearpar{liu2023itransformer}}} &
\multicolumn{2}{c|}{\scalebox{0.99}{\citeyearpar{zeng2023transformers}}} &
\multicolumn{2}{c|}{\scalebox{0.99}{\citeyearpar{nie2022time}}} & \multicolumn{2}{c|}{\scalebox{0.99}{\citeyearpar{wu2022timesnet}}} & \multicolumn{2}{c|}{\scalebox{0.99}{\citeyearpar{zhou2022fedformer}}} & \multicolumn{2}{c|}{\scalebox{0.99}{\citeyearpar{wu2021autoformer}}} & \multicolumn{2}{c|}{\scalebox{0.99}{\citeyearpar{zhang2022less}}}  & \multicolumn{2}{c|}{\scalebox{0.99}{\citeyearpar{zhou2021informer}}} & \multicolumn{2}{c}{\scalebox{0.99}{\citeyearpar{kitaev2020reformer}}}  \\

\midrule

\multicolumn{1}{c|}{Metric} & MSE  & MAE & MSE & MAE& MSE & MAE& MSE  & MAE& MSE  & MAE& MSE  & MAE& MSE  & MAE& MSE  & MAE& MSE  & MAE& MSE  & MAE\\
\midrule
\multirow{1}{*}{\rotatebox{0}{\textit{UAVu1}}}
&\boldres{0.139}&\boldres{0.344}&\secondres{0.168}&\secondres{0.394}&0.237&0.552&0.185&0.467&0.289&0.599&0.349&0.618&0.432&0.638&0.427&0.699&0.627&0.812&0.663&0.912\\
\midrule

\multirow{1}{*}{\rotatebox{0}{\textit{UAVu2}}}
&\boldres{0.119}&\boldres{0.142}&\secondres{0.128}&\secondres{0.189} &0.204&0.329&0.190&0.296&0.301&0.379&0.391&0.399&0.401&0.426&0.508&0.465&0.603&0.635&0.699&0.613\\
\midrule

\multirow{1}{*}{\rotatebox{0}{\textit{ETTh1}}}
&\boldres{0.412} &\boldres{0.427} &0.465&0.455&0.422&0.437&\secondres{0.413}&\secondres{0.430}&0.458&0.450&0.440&0.460&0.496&0.487&0.491&0.479&1.040&0.795&1.029&0.805\\
\midrule

\multirow{1}{*}{\rotatebox{0}{\textit{ETTh2}}}
&\boldres{0.324}&\boldres{0.371}&0.381&0.412&0.431&0.446&\secondres{0.330}&\secondres{0.379}&0.414&0.427&0.437&0.449&0.450&0.459&0.602&0.543&4.431&1.729&6.736&2.191\\
\midrule

\multirow{1}{*}{\rotatebox{0}{\textit{ETTm1}}}
&\boldres{0.331}&\boldres{0.377}&0.388&0.403&0.357&\secondres{0.378}&\secondres{0.351}&0.380&0.400&0.406&0.448&0.452&0.588&0.517&0.435&0.437&0.961&0.734&0.799&0.671 \\
\midrule

\multirow{1}{*}{\rotatebox{0}{\textit{ETTm2}}}
&\secondres{0.256}&\boldres{0.315}&0.284&0.339&0.267&0.333&\boldres{0.255}&\secondres{0.315}&0.291&0.333&0.305&0.349&0.327&0.371&0.409&0.436&1.410&0.810&1.479&0.915 \\

\bottomrule
\end{tabular}
}
\end{center}
\end{table}





\vspace{-1em}
\section{Conclusion}
In this paper, we propose UDuo, a general-purpose online allocation dual optimization framework that achieves transferable solutions through time-series forecasting of our introduced user arrival representation vectors. By theoretically proving its constraints on feasibility guarantees, UDuo overcomes the limitations of conventional stochastic arrival model-based approaches. To further enhance the framework's generalizability, we incorporate a budget pacing module enabling real-time adjustment control planning, while the designed MiFormer architecture strengthens time-series forecasting capabilities for diverse response models and traffic patterns. Our experimental results demonstrate UDuo's superior performance in real-world online matching scenarios. Additionally, MiFormer validates its multi-scenario capabilities on benchmark tests and production data. This work establishes a foundational methodology for the prediction of future user distribution in online matching tasks. Future directions include extending UDuo to multi-constraint optimization scenarios with theoretical superiority verification, as well as integrating Large Time-series Forecasting models (LTM) to improve prediction accuracy for few-shot and zero-shot user arrival scenarios.






\bibliographystyle{unsrtnat}
\bibliography{mybibfile}







\appendix


\section{Implementation Details}
\label{implement}
\paragraph{Pacing Strategy.}We support two pacing strategies: timing-based pacing and generative pacing, designed to flexibly allocate the total daily budget across temporal segments while controlling expenditure rates. To isolate and validate the effectiveness of our temporal forecasting solver module, we adopt the simpler timing-based pacing strategy in experiments, distributing budgets across time slots according to historical statistical patterns.

\paragraph{Hyperparameters.}The temporal model is trained with the batch size 1024 using the Adagrad optimizer while the learning rate is set 0.012, where hidden states are set to 512 dimensions. The MiFormer architecture adopts a 4-layer encoder and 1-layer decoder configuration. The hyperparameters are selected via grid search on validation sets with early stopping (patience=3 epochs). Model training is conducted on NVIDIA A800-80GB GPUs, while online inference experiments are evaluated on A10 GPUs to ensure realistic deployment conditions.

\section{Experiment Details.}
\label{exp_setting}
\paragraph{Datasets Description.} We collected user traffic data from two major cities in Elema and employed distinct uplift models to generate response scores. Based on Eq.~\ref{vlambda}, the user arrival representation vectors (UAV) were aggregated into temporal sequences, resulting in two datasets: UAVu1 and UAVu2. Additionally, we evaluated long-term forecasting performance across six datasets, including four well-established benchmarks (ETTh1, ETTh2, ETTm1, ETTm2) and our production-scale datasets. Detailed dataset descriptions are provided in Table~\ref{datasets}.

\begin{table}[h]
\renewcommand\arraystretch{1.1}
\setlength{\tabcolsep}{3.5pt} 
\caption{Detailed dataset descriptions. Dataset sizes are listed as (Train, Validation, Test).}
\label{datasets}
\vspace{-3mm}
\begin{center}
\begin{scriptsize} 
\begin{tabular}{@{}ccccccc@{}}
\toprule
Tasks & Dataset & Dim & Series Length & Dataset Size & Frequency & Information \\ \midrule
\multirow{6}{*}{\parbox{3.8cm}{\centering Long-term Forecasting}} 
& UAVu1 & 9 & \{12,24,48,96\} 
& (3,514,678, 479,154, 959,308)
& 15min & Users \\
& UAVu2 & 9 & \{12,24,48,96\} 
& (3,124,471, 399,341, 912,303)
& 15min & Users \\
& ETTm1 & 7 & \{96,192,336,720\} 
& (34,465, 11,521, 11,521) 
& 15min & Temperature \\
& ETTm2 & 7 & \{96,192,336,720\} 
& (34,465, 11,521, 11,521) 
& 15min & Temperature \\
& ETTh1 & 7 & \{96,192,336,720\} 
& (8,545, 2,881, 2,881) 
& Hourly & Temperature \\
& ETTh2 & 7 & \{96,192,336,720\} 
& (8,545, 2,881, 2,881) 
& Hourly & Temperature \\ \bottomrule
\end{tabular}
\end{scriptsize}
\end{center}
\end{table}

\paragraph{Evaluation Metrics.} To evaluate the performance of our temporal forecasting model, we conduct quantitative assessments from the perspective of predictive accuracy. Evaluation metrics include Mean Squared Error (MSE) \cite{hyndman2006another} and Mean Absolute Error (MAE) \cite{willmott2005advantages} for long-term forecasting tasks. The mathematical formulations of these metrics are defined as follows:
\begin{equation}
    MSE=\frac{1}{T}\sum_{t=1}^{T}(y_t-\hat{y_t})^2
\end{equation}
\begin{equation}
    MAE=\frac{1}{T}\sum_{t=1}^{T}|y_t-\hat{y_t}|
\end{equation}
where $y$,$\hat{y}\in \mathbb{R}^{T\times C}$ are the ground truth and forecasting results of the future with $L$ time points and $C$ dimensions. $y_t$ means the $t$-th future time point.

\paragraph{Baselines.} To validate the effectiveness of our proposed method, we evaluate MiFormer against 9 baselines representing the state-of-the-art in long-term forecasting domains. These benchmarks are applied to in-domain task-specific full-sample prediction assessments, including iTransformer\cite{liu2023itransformer}, PatchTST\cite{zeng2023transformers}, DLinear \cite{nie2022time}, TimesNet\cite{wu2022timesnet}, FEDformer\cite{zhou2022fedformer}, Autoformer\cite{wu2021autoformer}, LightTS\cite{zhang2022less}, Informer\cite{zhou2021informer}, and Reformer\cite{kitaev2020reformer}.



\newpage

\section*{NeurIPS Paper Checklist}

The checklist is designed to encourage best practices for responsible machine learning research, addressing issues of reproducibility, transparency, research ethics, and societal impact. Do not remove the checklist: {\bf The papers not including the checklist will be desk rejected.} The checklist should follow the references and follow the (optional) supplemental material.  The checklist does NOT count towards the page
limit. 

Please read the checklist guidelines carefully for information on how to answer these questions. For each question in the checklist:
\begin{itemize}
    \item You should answer \answerYes{}, \answerNo{}, or \answerNA{}.
    \item \answerNA{} means either that the question is Not Applicable for that particular paper or the relevant information is Not Available.
    \item Please provide a short (1–2 sentence) justification right after your answer (even for NA). 
\end{itemize}

{\bf The checklist answers are an integral part of your paper submission.} They are visible to the reviewers, area chairs, senior area chairs, and ethics reviewers. You will be asked to also include it (after eventual revisions) with the final version of your paper, and its final version will be published with the paper.

The reviewers of your paper will be asked to use the checklist as one of the factors in their evaluation. While "\answerYes{}" is generally preferable to "\answerNo{}", it is perfectly acceptable to answer "\answerNo{}" provided a proper justification is given (e.g., "error bars are not reported because it would be too computationally expensive" or "we were unable to find the license for the dataset we used"). In general, answering "\answerNo{}" or "\answerNA{}" is not grounds for rejection. While the questions are phrased in a binary way, we acknowledge that the true answer is often more nuanced, so please just use your best judgment and write a justification to elaborate. All supporting evidence can appear either in the main paper or the supplemental material, provided in appendix. If you answer \answerYes{} to a question, in the justification please point to the section(s) where related material for the question can be found.

IMPORTANT, please:
\begin{itemize}
    \item {\bf Delete this instruction block, but keep the section heading ``NeurIPS Paper Checklist"},
    \item  {\bf Keep the checklist subsection headings, questions/answers and guidelines below.}
    \item {\bf Do not modify the questions and only use the provided macros for your answers}.
\end{itemize}


\begin{enumerate}

\item {\bf Claims}
    \item[] Question: Do the main claims made in the abstract and introduction accurately reflect the paper's contributions and scope?
    \item[] Answer: \answerTODO{} 
    \item[] Justification: \justificationTODO{}

\item {\bf Limitations}
    \item[] Question: Does the paper discuss the limitations of the work performed by the authors?
    \item[] Answer: \answerTODO{} 
    \item[] Justification: \justificationTODO{}

\item {\bf Theory assumptions and proofs}
    \item[] Question: For each theoretical result, does the paper provide the full set of assumptions and a complete (and correct) proof?
    \item[] Answer: \answerTODO{} 
    \item[] Justification: \justificationTODO{}

    \item {\bf Experimental result reproducibility}
    \item[] Question: Does the paper fully disclose all the information needed to reproduce the main experimental results of the paper to the extent that it affects the main claims and/or conclusions of the paper (regardless of whether the code and data are provided or not)?
    \item[] Answer: \answerTODO{} 
    \item[] Justification: \justificationTODO{}

\item {\bf Open access to data and code}
    \item[] Question: Does the paper provide open access to the data and code, with sufficient instructions to faithfully reproduce the main experimental results, as described in supplemental material?
    \item[] Answer: \answerTODO{} 
    \item[] Justification: \justificationTODO{}

\item {\bf Experimental setting/details}
    \item[] Question: Does the paper specify all the training and test details (e.g., data splits, hyperparameters, how they were chosen, type of optimizer, etc.) necessary to understand the results?
    \item[] Answer: \answerTODO{} 
    \item[] Justification: \justificationTODO{}

\item {\bf Experiment statistical significance}
    \item[] Question: Does the paper report error bars suitably and correctly defined or other appropriate information about the statistical significance of the experiments?
    \item[] Answer: \answerTODO{} 
    \item[] Justification: \justificationTODO{}

\item {\bf Experiments compute resources}
    \item[] Question: For each experiment, does the paper provide sufficient information on the computer resources (type of compute workers, memory, time of execution) needed to reproduce the experiments?
    \item[] Answer: \answerTODO{} 
    \item[] Justification: \justificationTODO{}
    
\item {\bf Code of ethics}
    \item[] Question: Does the research conducted in the paper conform, in every respect, with the NeurIPS Code of Ethics \url{https://neurips.cc/public/EthicsGuidelines}?
    \item[] Answer: \answerTODO{} 
    \item[] Justification: \justificationTODO{}

\item {\bf Broader impacts}
    \item[] Question: Does the paper discuss both potential positive societal impacts and negative societal impacts of the work performed?
    \item[] Answer: \answerTODO{} 
    \item[] Justification: \justificationTODO{}
    
\item {\bf Safeguards}
    \item[] Question: Does the paper describe safeguards that have been put in place for responsible release of data or models that have a high risk for misuse (e.g., pretrained language models, image generators, or scraped datasets)?
    \item[] Answer: \answerTODO{} 
    \item[] Justification: \justificationTODO{}

\item {\bf Licenses for existing assets}
    \item[] Question: Are the creators or original owners of assets (e.g., code, data, models), used in the paper, properly credited and are the license and terms of use explicitly mentioned and properly respected?
    \item[] Answer: \answerTODO{} 
    \item[] Justification: \justificationTODO{}

\item {\bf New assets}
    \item[] Question: Are new assets introduced in the paper well documented and is the documentation provided alongside the assets?
    \item[] Answer: \answerTODO{} 
    \item[] Justification: \justificationTODO{}

\item {\bf Crowdsourcing and research with human subjects}
    \item[] Question: For crowdsourcing experiments and research with human subjects, does the paper include the full text of instructions given to participants and screenshots, if applicable, as well as details about compensation (if any)? 
    \item[] Answer: \answerTODO{} 
    \item[] Justification: \justificationTODO{}

\item {\bf Institutional review board (IRB) approvals or equivalent for research with human subjects}
    \item[] Question: Does the paper describe potential risks incurred by study participants, whether such risks were disclosed to the subjects, and whether Institutional Review Board (IRB) approvals (or an equivalent approval/review based on the requirements of your country or institution) were obtained?
    \item[] Answer: \answerTODO{} 
    \item[] Justification: \justificationTODO{}

\item {\bf Declaration of LLM usage}
    \item[] Question: Does the paper describe the usage of LLMs if it is an important, original, or non-standard component of the core methods in this research? Note that if the LLM is used only for writing, editing, or formatting purposes and does not impact the core methodology, scientific rigorousness, or originality of the research, declaration is not required.
    \item[] Answer: \answerTODO{} 
    \item[] Justification: \justificationTODO{}

\end{enumerate}

\end{document}